# Performance of Single User vs. Multiuser Modulation in Wireless Multicarrier (MC) Communications

Anwarul Azim, *Lecturer, East West University Bangladesh*

*Abstract--* **The main objective of this paper is to compare block transmission system performance analytically for Code Divisional Multiple Access (CDMA) modulations in Generalized Multicarrier environment against linear modulation techniques for both single user and multiuser. The effectivity of GMC-CDMA for multiuser will also be judged for different Direct Sequence CDMA (DS-CDMA) such as MC-CDMA, MC-DS-CDMA, MC-SS-CDMA. The analytical comparison will be in terms of computing probability of bit error for frequency selective and slow flat fading channels for different modulation techniques. The Bit Error Rate should be a good indication for of the performance. The tolerance characteristics of DS-CDMA in frequency-selective channels and MC-CDMA in flat fading channels will be shown analytically and improved capacity and Bit Error Rate performance will be derived for Block Spread Multiuser Multicarrier by relying block symbol spreading and Fast Fourier Transform (FFT) operations. GMC-CDMA should give guaranteed symbol recovery regardless of channel limitations.**

*Index terms*—Code Division Multiple Access (CDMA), Direct Sequence CDMA (DS-CDMA), Generalized Multicarrier CDMA (GMC-CDMA).

## I. Introduction

In digital wireless communication systems [1, Chapter 6], several factors comes in to consideration to choose the suitable digital modulation technique. A desired modulation technique produces low Bit Error Rates (BER) at low received Signal-to-Noise ratios (SNR), performs well in multipath and fading conditions, occupies a minimum of bandwidth (BW) and is easy and cost effective to implement. The tradeoffs between BER performances vs. bandwidth efficiency is always made in terms of signal fidelity at low power levels and data accommodation within a limited bandwidth .Linear modulation techniques such as M-ary keying ensure bandwidth efficiency to perform for both single user and multiuser demands where as in multiuser multiple access environment, Spread Spectrum (SS) systems shows more robustness against interference and multipath fading with many users can simultaneously same BW without significant interference. Performance of Multiuser Modulation techniques [1] hold great potential in signal processing challenges in various applications such as audio/video broadcasting, cable networks, modem design, multimedia services, mobile local area networks and future generation wideband cellular systems.

## II. Block spread Multicarrier Multiuser Communication

Wireless multicarrier (MC) systems consist of multiple complex exponentials as information carriers became effective after FFT implementation. MC combining with direct sequence code division multiple access (DS-CDMA) spread spectrum (SS) systems have produced various wideband cellular communication standards. These MC and DS-CDMA systems transmit information in blocks of symbols processed through inverse fast fourier transform (IFFT) and they both faces inter block interference (IBI) because of impulse response of multipath channels. Redundancy added to each transmission block to eliminate the effects of IBI at the receiver. MC and SS transmissions offer comparable features to deal with the effects of frequency selective multipath fading channels causing inter symbol interference (ISI), multiuser interference (MUI) from high power near users masking user powers located far-away and Rayleigh fading channels.

As MC systems are has the orthogonal features through linear time invariant (LTI) implementation, it mitigates the effect of MUI. But DS-CDMA systems require multipath detection aiding MUI in wireless communication environment. But as an SS system, DS-CDMA spreads the information across the bandwidth (BW) which provides tolerance to multipath and frequency selective fading. Now to improve the capacity and bit error rate (BER) performance by reducing the effects of MUI and fading, MC and DS-CDMA can be combined into a generalized MC-CDMA (GMC-CDMA) communication system. The GMC-CDMA will be MUI and ISI resilient and has high probability of symbol recovery at the receiver even in frequency selective multipath fading environment with ensuring BW efficiency.

Transmission in wireless environment, channel induced ISI effects the system performance severely. To alleviate ISI which also results in frequency selective fading issues, information bearing chips will be transmitted in blocks.

A. *Generalized Multicarrier Code Divisional Multiple Access (GMC-CDMA)*

To build MUI and ISI tolerant transceiver, let us suppose there are $M$ users using the channels using impulse response for user l,

$$H_l(z) = z^{-d_l} \sum_{N=0}^{L} h_l(n) z^{-N} \qquad (1)$$

where $L$ is the max. delay spread of the chip sampled multipath channels. GMC-CDMA employs the frequency divisional multiple access (FDMA) principles by transmitting on channels with non-overlapping frequencies. It will also load the same symbol to more than $L$ subcarriers that results in at least one subcarrier being received in adverse channel conditions. This symbol recovery without BW over-expansion is called symbol blocking. The idea is to let each user send $K \geq L$ symbols using $J \geq K$ subcarriers instead of individual placement in separate subcarrier. The users' subcarriers would have FDMA standard separation in frequency domain. To eliminate MUI and ISI, each user block of $K$ symbols is precoded to $J$ symbols placed on $J$ subcarriers so that all K symbols in the block are recoverable from any $J$-$L$ of $J$ subcarriers. For $M$ users system, total number of subcarriers is required is $N=MJ$ and transmitting chips should have blocks of $N$.

Now for the block transmission design, the transmitted sequences are grouped in blocks of size $P = N+1$ that include $L$ length of cyclic prefixing or zero padding to cancel out IBI. The $l$ th user's channel will be represented as an $N$ X $N$ circulant matrix $H_l$ [Fig.1]. So user $l$'s transmitting N X 1 vector would be $u_l(n) = C_l s_l(n)$ where $C_l$ is defined as $N$ X $K$ spreading matrix for user $l$. So received vector, $x(n) = \sum_{l=0}^{M-1} H_l C_l s_l(n) + \eta(n)$ and user's block-symbol estimate, $s_l^*(n) = G_l x(n)$ where $G_l$ is the $K$ X $N$ receiving matrix for user $l$. This general system model can describe the principle operation of various DS-CDMA standards including combinations of MC and DS systems such as MC-CDMA, MC-DS-CDMA, MT-CDMA, and MC-SS-MA. This GMC-CDMA is depicted in figure 1.

Also to present the MUI/ISI resilience of the model, let us consider $C_l = F^H \alpha_l \Omega_l$ and $G_l = \Gamma_l \alpha_l^T F$ for spreading and de-spreading matrices. Here $\Omega_l$ is a $J$ X $K$ matrix that linearly maps the $K$ information symbols of the nth block $s_l(n)$ to $J$ symbols $\Omega_l s_l(n)$ through $N$ X $J$ subcarrier matrix $\alpha_l$. The IFFT matrix $F^H$ implements an OFDM modulation at the final stage. At the receiver end, $\Gamma_l$ equalizes the channel precoder combination while F represents the FFT operation on $x(n)$ where $\alpha_l$ decodes J symbols from subcarriers for user $l$.

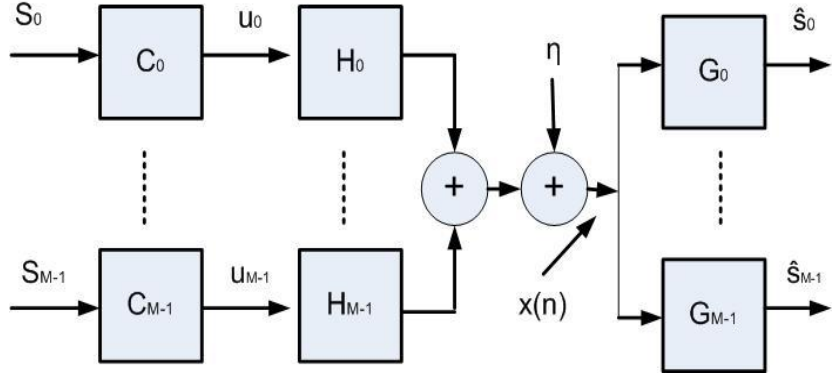

Figure 1. Block Spreading System

So the user's symbol estimate can be re-written as,

$$s_l^*(n) = \Gamma_l \sum_{q=0}^{M-1} \alpha_l^T F H_q F^H \alpha_q \Omega_q s_q(n) + \Gamma_l \alpha_L^T F \eta(n) \quad (2)$$

Using the properties from [3], we can eliminate the effect of MUI and express the estimate as,

$$s_l^*(n) = \Gamma_l \lfloor D_l \Omega_l s_l(n) + \alpha_l F \eta(n) \rfloor = \Gamma_l x_l(n) \quad (3)$$

where $x_l(n)$ is the MUI-free vector for user $l$. Considering the associated noise terms are white Gaussian we can conclude that this GMC-CDMA design delivered the multipath equivalent to independent parallel single user frequency selective channels with AWGN. This model performs better with symbol detectability with robustness to frequency fading.

For other design considerations such as block length $K$, the number of redundant symbols $L$ and the number of users $M$, it has to be done in terms of BW utilization. If each user sends $K$ information symbols for $P=N+L=MJ+L$ transmitted symbols. So the bandwidth (BW) efficiency should be, $\lambda = \dfrac{MK}{P} = \dfrac{MK}{M(K+1)+L} \leq 1$. So larger $K$ will give higher BW efficiency which will be affecting system performance.

GMC-CDMA also has the capability to assign variable transmission rates to different users. As a result, users have different number of subcarriers allocated for their symbol transmission. This multirate feature will enable the transmission scheme to rate switching, better Rate resolution and in general better BER performance [5]. This multirate service is incorporated while preserving MUI and ISI resilience.

A. *Multirate Capabilities of GMC-CDMA*

Multirate systems are attractive for their variable rates for different services such as text, images, data and low rate video which can meet different Quality of Service (QoS) requirements. DS-CDMA systems have the design flexibility and improved capacity through multicarrier to provide multirate services by choosing appropriate chip rate, variable spreading length, number of multiple codes and modulation methods. The receiver design for DS-CDMA systems may vary from maximum likelihood (ML) decoders, conventional matched filters (MF) or de-correlating multichannel equalizers, minimum mean square error (MMSE) receivers, successive interference cancellers, decision feedback receivers. But in multipath environment, these designs fail to mitigate MUI and ISI effects without high complexity and may also cause noise enhancement. The proposed GMC-CDMA gives a receiver design which can be MUI/ISI free and it can guarantee symbol recovery in presence frequency selective multipath fading and Rayleigh fading. Such a system is shown in figure 2 [5].

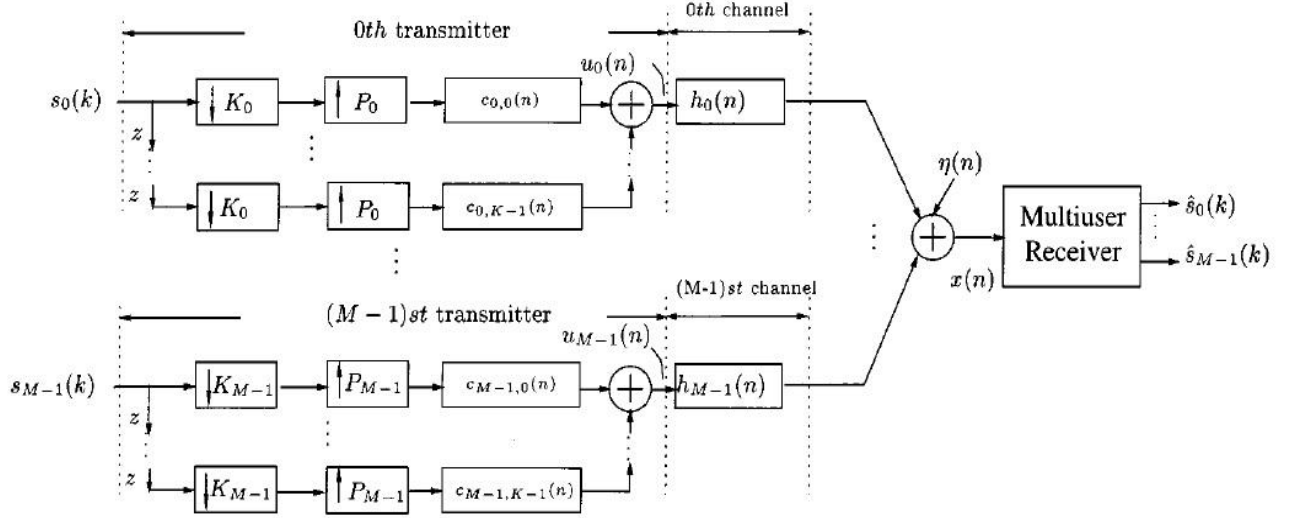

Figure 2. Multirate Block Precoded CDMA System

Now for user $l$, a block of $K_l$ symbols is transmitted using $P_l$ chips; information rate for the user would be $R_l = \dfrac{K_l}{P_l T_c}$ symbols/second where $T_C$ the chip/sampling period is. So the total rate $R_T$ would be,

$$R_T = \sum_{l=0}^{M-1} R_l = \frac{MK}{T_c P} = \frac{M(J-L)}{T_c(MJ+L)} \qquad (4)$$

For multirate design, $R_T$ has to come close to the available BW $1/T_c$ with a decoding delay. After required rate adjustment and code design, GMC-CDMA converts a frequency selective channel to a set of flat fading channels and linearly precoded symbols on the subcarriers. As a result, each symbol is spread on multiple subcarriers achieving frequency diversity mitigating fading effect.

The main advantage of the GMC-CDMA is that it uses multipath diversity which comes from the block-spreading to enable MUI/ISI elimination by design. Now if the symbols of $S_l$ are binary and iid, the BER performance of the $k$th symbol in terms of generic vector $v$ and moment generating functions of the symbol, ISI and noise would be [5],

$$P_{e,l,k} = \frac{1}{2\pi j} \int v^{-1} \gamma_{l,k}^{symbol}(v) \gamma_{l,k}^{ISI}(v) \gamma_{l,k}^{noise}(v) dv \qquad (5)$$

This BER performance is to ensure MUI/ISI resilient transmission with guaranteed channel identification ability and symbol recovery at the receiver in frequency selective multipath environment.

III. CONCLUSION

The proposed GMC-CDMA model relies on symbol blocking and blocks precoding but it make decoding computationally simple. Its multirate receiver design performs better than existing frameworks in fading conditions through MUI elimination, blind channel estimation and FIR channel irrespective symbol recovery with finer rate resolution and easier rate switching capabilities. It exploits the length of block spreading codes at the expense of longer delays in decoding at the receiver side. But reduced complexity and increased flexibility in the design makes GMC-CDMA ideal for robust MUI/ISI free multipath multicarrier communication.